\documentclass[3p,times,procedia]{elsarticle}
\usepackage{nupha_ecrc}

\volume{00}

\firstpage{1}

\journalname{Nuclear Physics A}

\runauth{}

\jid{nupha}

\jnltitlelogo{Nuclear Physics A}

\usepackage{amssymb}
\usepackage[figuresright]{rotating}

\begin{document}

\begin{frontmatter}



%% Instructions from Editor: Please use the following \dochead only in the preprint version (e-print arXiv etc.); 
%% use empty \dochead{} when submitting to Nuclear Physics A!
\dochead{XXVIth International Conference on Ultrarelativistic Nucleus-Nucleus Collisions\\ (Quark Matter 2017)}
%% e.g. \dochead{17th International Conference on Dynamical Processes in Excited States of Solids}

\title{Electromagnetic fields from quantum sources in heavy-ion collisions}


\author{B. Peroutka and K. Tuchin}

\address{Department of Physics and Astronomy, Iowa State University, Ames, Iowa, 50011, USA}

\begin{abstract}

We compute the electromagnetic field created by an ultrarelativistic charged  particle in vacuum at distances comparable to the particle Compton wavelength. The wave function of the particle is governed by the Klein-Gordon equation, for a scalar particle, or the Dirac equation, for a spin-half particle. The produced electromagnetic field is essentially different in magnitude and direction from the Coulomb field, induced by a classical point charge, due to the quantum diffusion effect. Thus, a realistic computation of the electromagnetic field produced in heavy-ion collisions must be based upon the full quantum treatment of the valence quarks.

\end{abstract}


\end{frontmatter}

% \linenumbers

\section{Electromagnetic field of classical charges}\label{sec1}

\begin{figure}[ht]
\begin{tabular}{cc}
      \includegraphics[height=5cm]{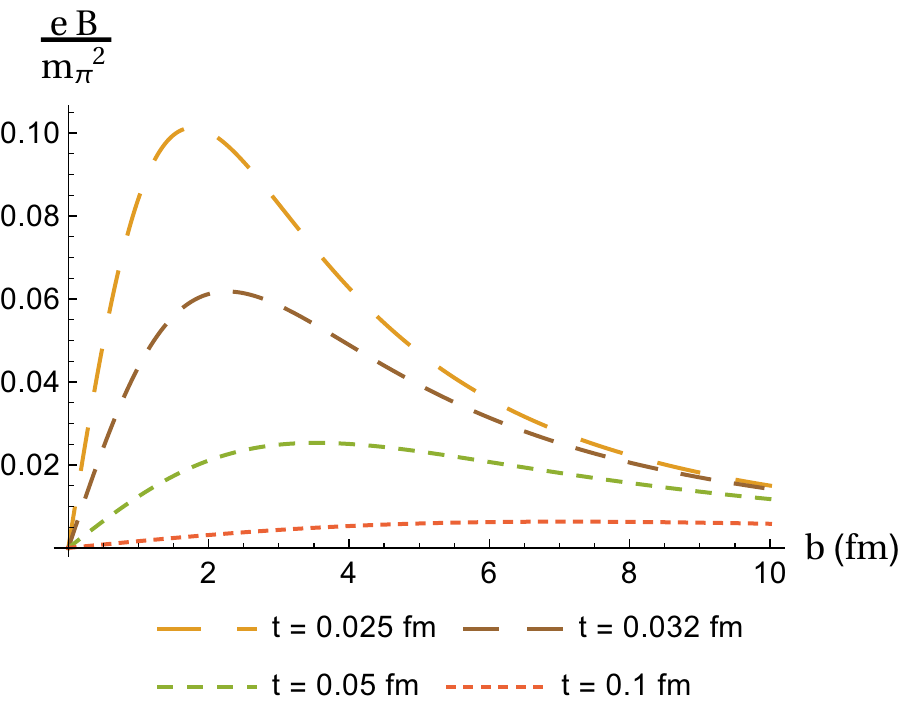} &
      \includegraphics[height=5cm]{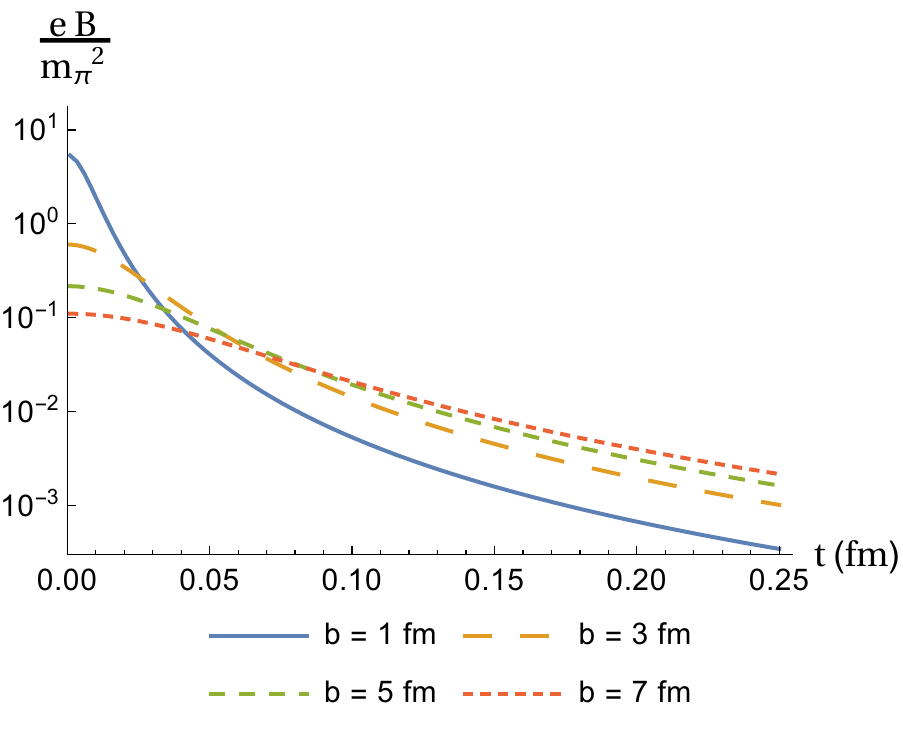}
      \end{tabular}
  \caption{Magnetic field $B$ created by a single classical point charge moving along the $z$-axis with velocity $v$ in vacuum as a function of impact parameter $b$ (left panel) and time $t$ (right panel),  see Eq.~(\ref{b15}).  }
\label{Classical-B}
\end{figure}

An intense electromagnetic field is produced in high energy hadron and nuclear collisions by the spectator valence quarks. It can have a significant phenomenological impact especially in heavy-ion collisions. There are many approaches to compute the field strength \cite{Kharzeev:2007jp,Skokov:2009qp,Voronyuk:2011jd}, but they all rely on the classical approximation that treats the valence quarks as point particles. The corresponding field is given by 
\begin{equation}\label{b15}
{\bf B}=  \frac{\gamma e v \hat {\bf \phi}}{4\pi}\frac{b}{(b^2+\gamma^2(vt-z)^2)^{3/2}}\,,\quad 
{\bf E}=  \frac{\gamma e }{4\pi}\frac{{\bf b} + (z-vt)\hat{\bf z}}{(b^2+\gamma^2(vt-z)^2)^{3/2}}\,,
\end{equation}
and shown in Fig.~\ref{Classical-B}. This is a good approximation at large distances, where the multipole expansion is valid; however, in a realistic heavy-ion collision, the interaction range, the quark wave function size, and the dimensions of the produced nuclear matter all have similar extent. Therefore, one has to treat the valance quark current quantum mechanically.

%%%%
%%%%%

\section{Electromagnetic field of  quantum scalar charges}\label{sec2}

\begin{figure}[ht]
\begin{tabular}{cc}
      \includegraphics[height=5cm]{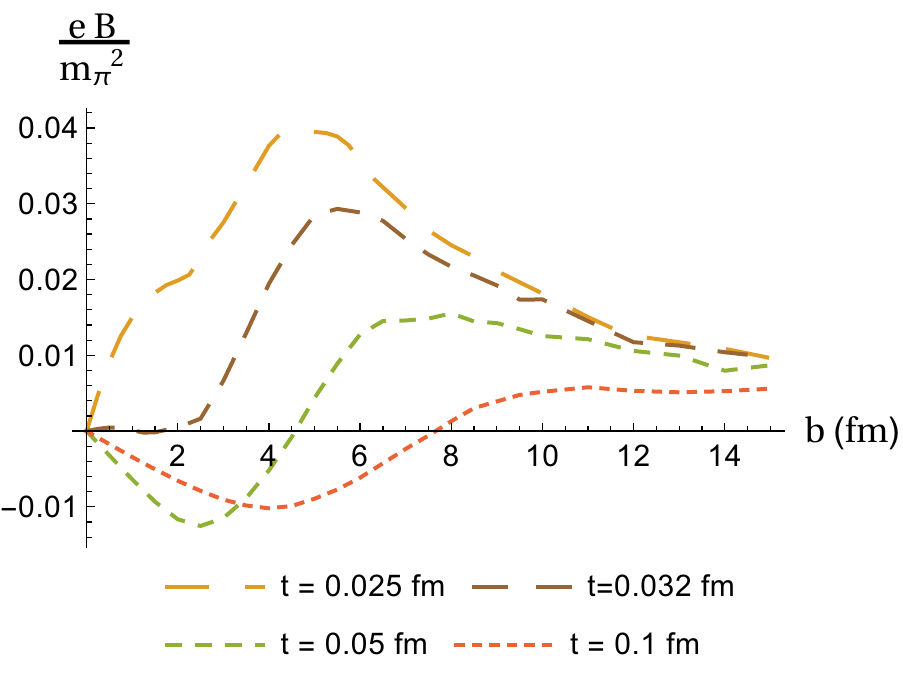} &
      \includegraphics[height=5cm]{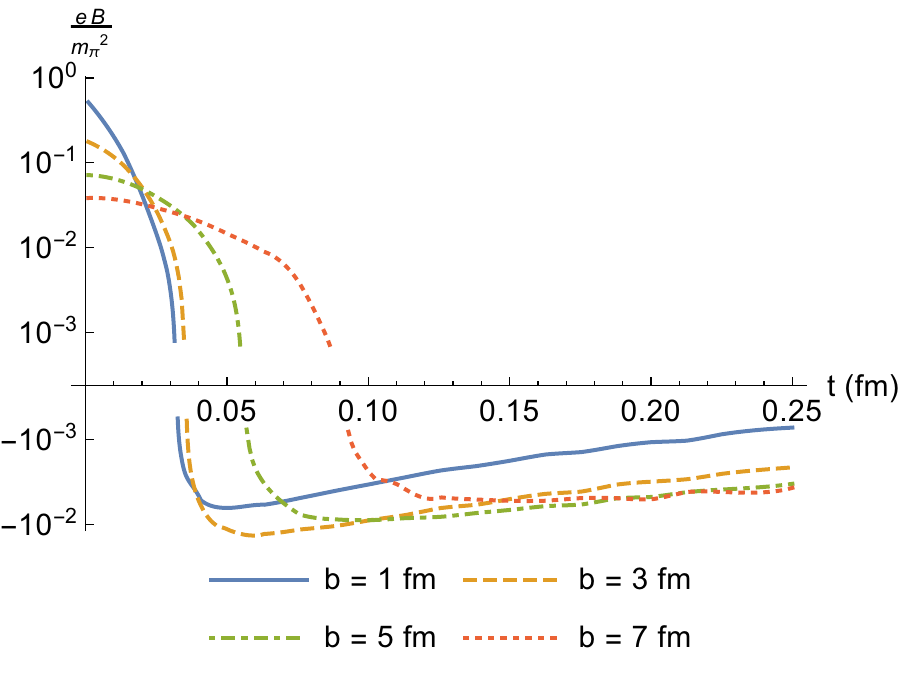}
      \end{tabular}
  \caption{Magnetic field  $B$ created by a quantum charge in vacuum vs impact parameter $b$ (left panel) and time $t$ (right panel). $a=1$~fm, $m=0.3$~GeV,  $\gamma=100$.
}
\label{Quantum-B}
\end{figure}

As the first step towards understanding the quantum dynamics of the electromagnetic field sources,  we model the valence quarks as spinless Gaussian wave packets \cite{Holliday:2016lbx}. The corresponding wave function in momentum space in the particle rest frame reads
\begin{equation}\label{d1}
\psi_{\bf k}(0)= \left(\frac{a^2}{\pi \hbar^2}\right)^{3/4}
e^{-\frac{a^2k^2}{2\hbar^2}}\,.
\end{equation}
The width of the packet $a$ is the only parameter describing the wave function. In all calculations, it is set to be 1~fm. Time-evolution of this packet is governed by the Klein-Gordon equation
\begin{equation}
\psi({\bf r}_0, t_0)= \int \frac{d^3k}{(2\pi \hbar)^{3/2}} e^{\frac{i{\bf k}\cdot {\bf r}_0}{\hbar}}e^{-\frac{i\epsilon_k t_0}{\hbar }}\, \sqrt{\frac{m}{\epsilon_k}}\psi_{\bf k}(0)\,,
\end{equation}
where $\epsilon_k= \sqrt{m^2+k^2}$. The subscript "0" indicates the particle rest frame. The corresponding charge and current densities are computed according to
\begin{equation}\label{h10}
\rho_0 = \frac{ie\hbar}{2m}[\psi^* \partial_t\psi - (\partial_t\psi^*)\psi]\,,\quad 
{\bf j}_0= \frac{e\hbar}{2mi}[\psi^* {\bf\nabla}\psi - \psi {\bf \nabla}\psi^*]\,.
\end{equation}
The electromagnetic field in the rest frame has only one radial electric component 
\begin{equation}\label{g6E}
{\bf E}_0({\bf r}_0, t_0)= \int  \left\{ \frac{\rho_0({\bf r}',t') {\bf R}}{R^3}+\frac{{\bf R}}{R^2}\frac{\partial \rho_0({\bf r}',t') }{\partial t'}-\frac{1}{R}\frac{\partial {{\bf j}_0}({\bf r}',t')}{\partial t'}\right\} d^3r'= E_0(r_0,t_0)\hat {\bf r}\,,
\end{equation}
where ${\bf R}= {\bf r}-{\bf r}'$ and $t'=t - R$ is the retarded time. In the laboratory frame, where the charge is moving with velocity $v$ along the positive $z$ axis, the electromagnetic field is given by ${\bf E}= B \hat {\bf b}/v+E_z\hat {\bf z}$, $\bf{B}= B\hat {\bf\phi}$, where 
\begin{equation} \label{c13}
B({\bf r}, t)= E_0\left(\sqrt{b^2+\gamma^2(z-vt)^2},\gamma(t-vz)\right)\frac{v\gamma  b}{\sqrt{b^2+\gamma^2(z-vt)^2}}\,.
\end{equation}
The result of the numerical calculation is shown in Fig.~\ref{Quantum-B}. While at early times the classical and quantum calculations looks qualitatively similar, at later times, they are very different. The magnitude of the field of the quantum sources at $t=0.25$~fm is an order of magnitude larger than the classical one because of spreading of the quark wave function in the transverse direction due to the quantum diffusion. Furthermore, unlike the fields of the classical sources, all components of fields of the quantum sources change sign. This occurs because the quantum diffusion current increases while the charge density decreases with time at small impact parameters $b$ and large times $t$.

%%%%
%%%%%

\section{Electromagnetic field of quantum spin-1/2 charges}\label{sec3}

\begin{figure}[ht]
\begin{tabular}{cc}
      \includegraphics[height=5cm]{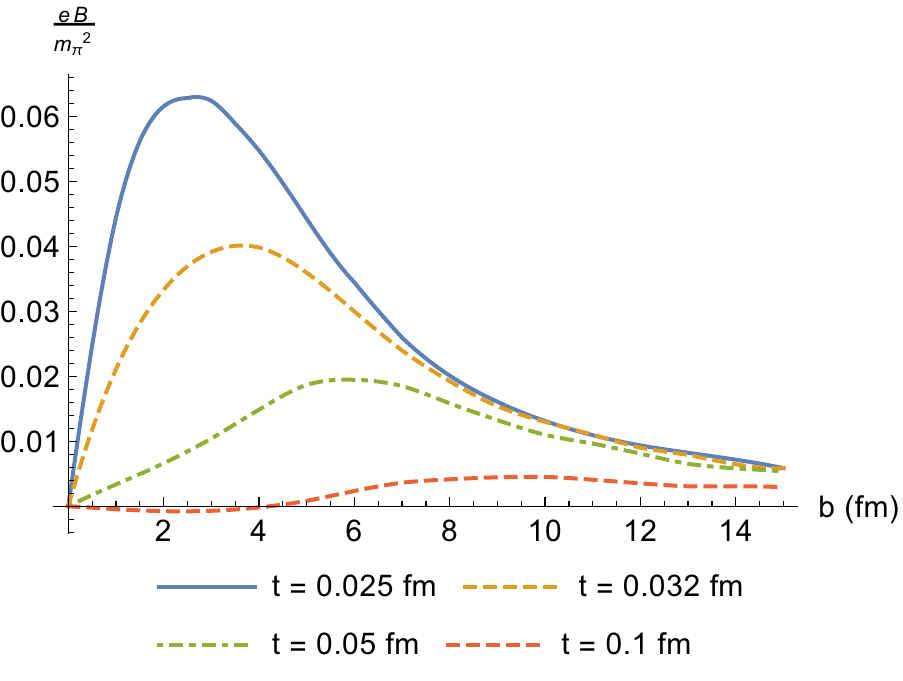} &
     \includegraphics[height=5cm]{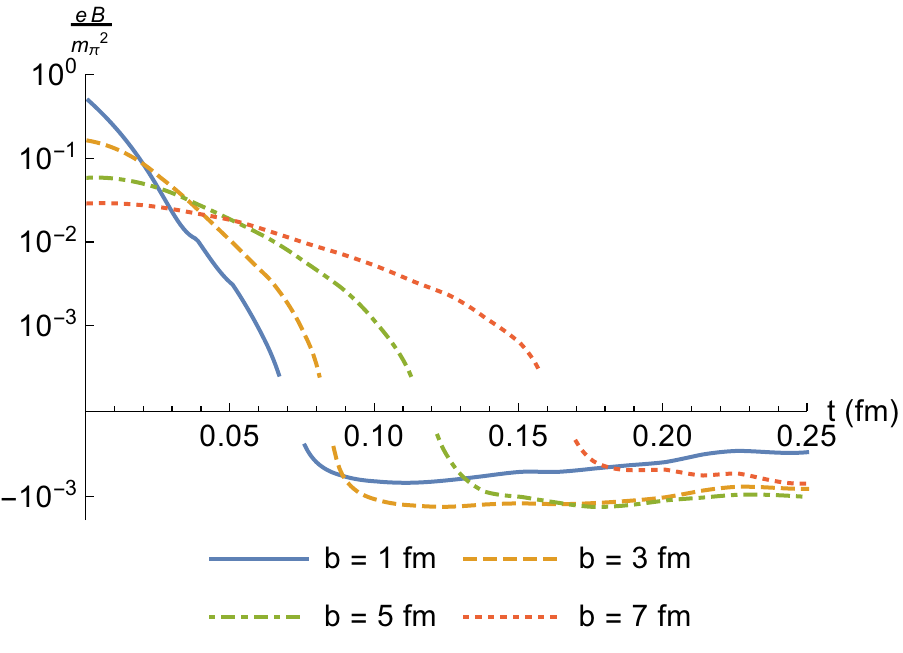}\\
       \end{tabular}
  \caption{ Magnetic field generated by a wave packet of  width $a=1$~fm in vacuum as a function of impact parameter $b$ (left panel) and time $t$ (right panel). Notice, that in the right panel, there is no discontinuity of the magnetic field, as might appear at first sight. It is an artifact of the logarithmic scale on the  vertical axis.}
\label{All}
\end{figure}

\begin{figure}[ht]
\begin{tabular}{cc}
      \includegraphics[height=5cm]{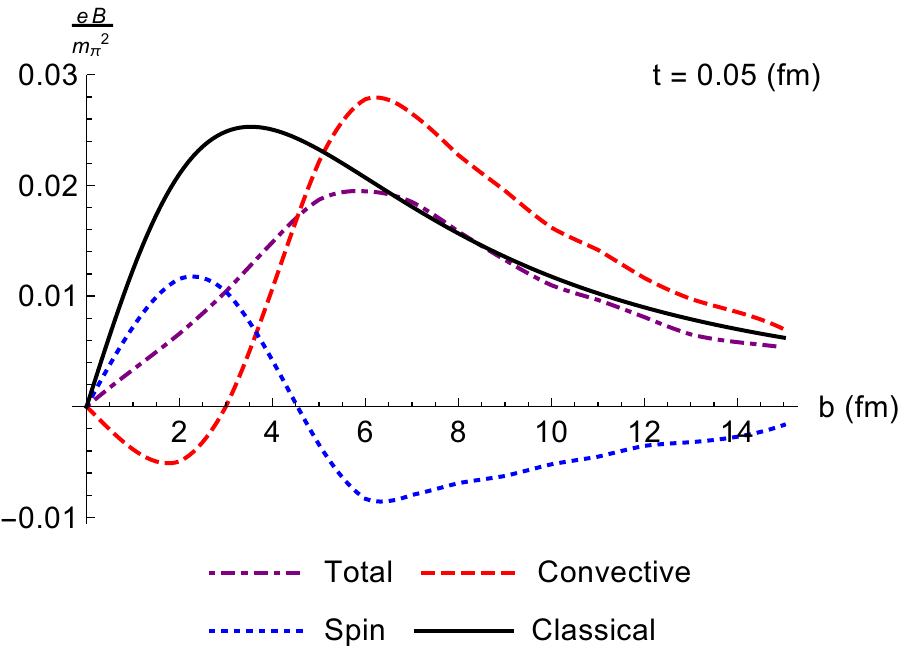} &
     \includegraphics[height=5cm]{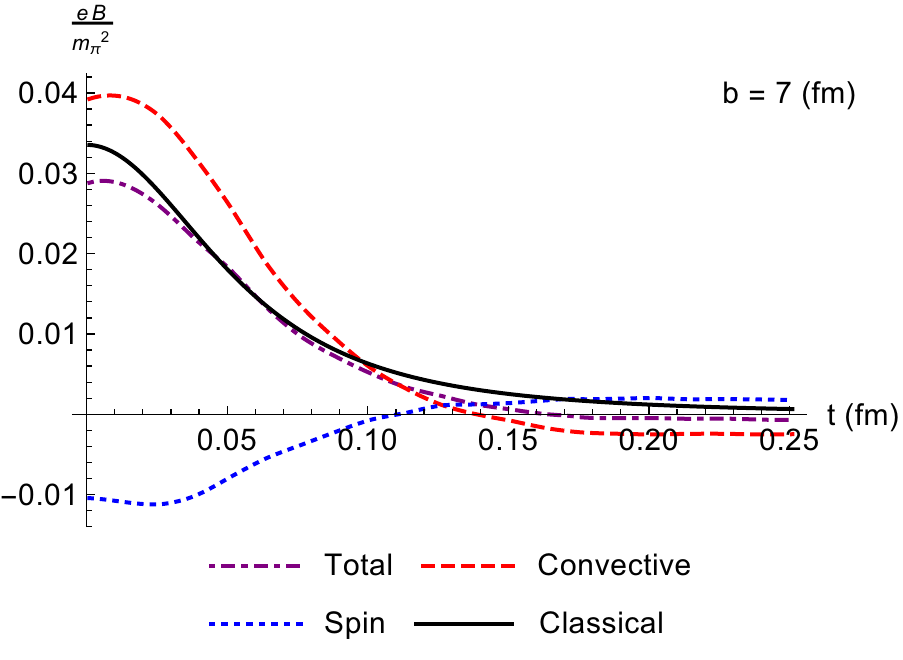}\\
       \end{tabular}
  \caption{Convective and spin current contributions to the magnetic field.}
\label{spin}
\end{figure}

In order to compute the effect of valence quark spin on the electromagnetic field, we describe the quark wave function by a bispinor
\begin{equation} \label{a5} 
\Psi({\bf r}, t) =\frac{1}{\sqrt{2}} \sum_{\lambda}\int \frac{d^3k}{(2\pi)^{3/2}}e^{i{\bf k} \cdot {\bf r}}e^{-i\epsilon_k t }\psi_{{\bf k}}(0)u_{{\bf k}\lambda}\,,
\end{equation}
where the four-component bispinor $u_{{\bf k}\lambda}$ is the momentum and helicity eigenstate given by 
\begin{equation} \label{a17}
u_{{\bf k} +}=  \sqrt{\frac{\epsilon_k+m}{2\epsilon_k}}\left(\begin{array}{c} \chi_+ \\  \frac{{\bf \sigma}\cdot {\bf k} }{\epsilon_k+m}\chi_+\end{array}\right)\,,
\quad 
u_{{\bf k} -}= \sqrt{\frac{\epsilon_k+m}{2\epsilon_k}} \left(\begin{array}{c} \chi_- \\  \frac{{\bf \sigma}\cdot {\bf k} }{\epsilon_k+m}\chi_-\end{array}\right)\,,
\end{equation}
where the two-component spinors $\chi_\pm$ are  helicity eigenstates. The corresponding charge and current densities read (in the rest frame)
\begin{equation} \label{b14}
\rho_0= e\Psi^\dagger \Psi\,,\quad {\bf j}_0 = e\Psi^\dagger{\bf \alpha} \Psi\,.
\end{equation}
Using the Gordon identity one can separate the convective and spin contributions. The electromagnetic field in the rest frame is computed according to Eq.~(\ref{g6E}). Then Eq.~(\ref{c13}) yields the magnetic field in the laboratory frame. The result of the numerical calculation is shown in Fig.~\ref{All} and Fig.~\ref{spin}. Although the magnetic field of spin-1/2 particles is qualitatively similar to that of scalar particles and is very different from the classical case, the sign-flip effect is less pronounced. This is because while the convective part of the magnetic field changes its sign from positive to negative, the spin part changes its sign from negative to positive at about the same time. One can see this in the right panel of Fig.~\ref{spin} where we use the linear scale for the vertical axis.

%%%%
%%%%%

In the left panel of Fig.~\ref{spin}  we plot each of the lines shown in Fig.~\ref{All} (left panel) separately along with its convective and spin components. Also plotted is the corresponding classical (boosted Coulomb) field for comparison. At large $b$, the classical (i.e.\ point) and quantum (i.e.\ wave packet) sources induce the same field as expected. While the convective current contributes to the monopole term falling off as $1/b^2$ at large $b$, the leading spin current contribution starts with the dipole term which falls off as $1/b^3$. At later times, due to the quantum diffusion, the deviation from the classical field is observed in a wider range of distances.

%%%%
%%%%%

\section{Conclusion}\label{sec4}

Describing the valence quark by a Gaussian wave packet, we computed the electromagnetic field it induces in vacuum and compared it to the electromagnetic field produced by the classical point charge. We observed that the classical  description of the valence quarks as point-particles is not accurate for calculations of the electromagnetic field in hadron and/or nuclear collisions as it breaks down at distances as large as $6$~fm at $\gamma=100$. Moreover, it misses an important  sign-flip effect that occurs due to the quantum diffusion of the wave packet. We conclude that a realistic computation of the electromagnetic field produced in heavy-ion collisions must be based upon full quantum treatment of the valence quarks.

\bigskip
  This work  was supported in part by the U.S. Department of Energy under Grant No.\ DE-FG02-87ER40371.



%% References


%\bibliographystyle{elsarticle-num}



\end{document}